\documentclass[iop]{emulateapj}

\usepackage{natbib,graphicx}
\begin{document}

\title{A Radio-Selected Black Hole X-ray Binary Candidate in the Milky Way Globular Cluster M62}
\author{Laura Chomiuk\altaffilmark{1,2}, Jay Strader\altaffilmark{2}, Thomas J.~Maccarone\altaffilmark{3}, James C.~A.~Miller-Jones\altaffilmark{4}, Craig Heinke\altaffilmark{5}, Eva Noyola\altaffilmark{6}, Anil C.~Seth\altaffilmark{7},  \& Scott Ransom\altaffilmark{1}}
\altaffiltext{1}{National Radio Astronomy Observatory, 520 Edgemont Rd, Charlottesville, VA, USA}
\altaffiltext{2}{Department of Physics and Astronomy, Michigan State University, East Lansing, Michigan 48824, USA}
\altaffiltext{3}{Department of Physics, Texas Tech University, Box 41051, Lubbock TX 79409-1051, USA}
\altaffiltext{4}{International Centre for Radio Astronomy Research, Curtin University, GPO Box U1987, Perth, WA 6845, Australia}
\altaffiltext{5}{Department of Physics, University of Alberta, 4-183 CCIS, Edmonton, AB T6G 2E1, Canada}
\altaffiltext{6}{Instituto de Astronomia, Universidad Nacional Autonoma de Mexico (UNAM), A.~P.~70-264, 04510, Mexico}
\altaffiltext{7}{Department of Physics and Astronomy, University of Utah, Salt Lake City, UT 84112, USA}

\email{chomiuk@pa.msu.edu}
\begin{abstract}
We report the discovery of a candidate stellar-mass black hole in the Milky Way globular cluster M62. We detected the black hole candidate, which we term M62-VLA1, in the core of the cluster using deep radio continuum imaging from the Karl G.~Jansky Very Large Array. M62-VLA1 is a faint source, with a flux density of $18.7\pm1.9$ $\mu$Jy at 6.2 GHz and a flat radio spectrum ($\alpha=-0.24\pm0.42$, for $S_{\nu} = \nu^{\alpha}$). M62 is the second Milky Way cluster with a candidate stellar-mass black hole; unlike the two candidate black holes previously found in the cluster M22, M62-VLA1 is associated with a \emph{Chandra} X-ray source, supporting its identification as a black hole X-ray binary. Measurements of its radio and X-ray luminosity, while not simultaneous, place M62-VLA1 squarely on the well-established radio--X-ray correlation for stellar-mass black holes. In archival \emph{Hubble Space Telescope} imaging, M62-VLA1 is coincident with a star near the lower red giant branch. This possible optical counterpart shows a blue excess, H$\alpha$ emission, and optical variability. The radio, X-ray, and optical properties of M62-VLA1 are very similar to those for V404~Cyg, one of the best-studied quiescent stellar-mass black holes. We cannot yet rule out alternative scenarios for the radio source, such as a flaring neutron star or background galaxy; future observations are necessary to determine whether M62-VLA1 is indeed an accreting stellar-mass black hole.

\end{abstract}

\section{Introduction}

For a typical globular cluster (GC) with a present day mass of $\sim$10$^{5}$--$10^{6}\ M_{\odot}$, hundreds of stellar-mass black holes (BHs) should be born during the first $\sim$10 Myr after formation \citep{Larson84, Kulkarni_etal93}. These BHs have masses $\sim$5--20 $M_{\odot}$ and mark the endpoints of stars with main-sequence masses $\gtrsim 25\ M_{\odot}$ \citep{Heger_etal03}. The specific frequency of X-ray binaries in GCs is $\sim$100 times larger than the field and provides strong evidence that mass-transferring binaries are dynamically formed with high efficiency in GCs \citep[e.g.,][]{Kundu_etal02, Pooley_etal03}. If BHs are present in GCs, a subset should be detectable as accreting binaries.

Yet, soon after the detection of luminous X-ray sources in Milky Way GCs, it became clear that accreting neutron stars, not BHs, dominate luminous cluster X-ray binaries.  The principal evidence is the detection of Type I X-ray bursts that are well-explained by thermonuclear runaway events on the surfaces of neutron stars \citep{Lewin_Joss81}.  Of the 18 luminous Galactic GC X-ray sources published so far, all are thought to contain neutron stars, due to the presence of X-ray bursts \citep[in 15; see e.g.][]{Verbunt_Lewin06,Altamirano_etal12}, coherent pulsations \citep[in 4; e.g.][]{Altamirano_etal10}, optical spectroscopy \citep{vanzyl04}, or a low radio-to-X-ray flux ratio \citep[][see below for the rationale]{Bozzo_etal11}.

Theoretical expectations for the presence of BHs in GCs remain uncertain. After birth, assuming the BHs receive small natal kicks, they will rapidly mass segregate to the cluster center in a fraction of a relaxation time and form a subcluster that is dynamically segregated from the rest of the GC. Many BH--BH binaries will then be formed through three-body interactions, and these binaries halt the collapse of the subcluster through interactions with one another. This process will tend to harden the binaries and lead to interactions with large recoil velocities, ejecting many BHs from the GC. Once a sufficient number of BHs have been ejected, the subcluster will no longer be dynamically decoupled from the stars in the core of the cluster, and the efficiency of BH ejection will decrease.

The eventual fate of the remaining BHs is poorly understood. Many papers have suggested that GCs should have at most one BH (or BH--BH binary) remaining at the present day, and that many clusters should have none \citep{Kulkarni_etal93, Sigurdsson_Hernquist93, pzm, Kalogera_etal04}. Some recent papers have argued a few BHs may be retained \citep{Mackey_etal08, Moody_Sigurdsson09, Aarseth12}, while two new theory papers suggest that in some circumstances many BHs could survive in GCs \citep{Sippel_Hurley13, Morscher_etal13}. In sum, recent work supports the idea that BH ejection in GCs is less efficient than previously thought.

\begin{figure*}
\begin{center}
\vspace{-1.0cm}
\includegraphics[width=8cm]{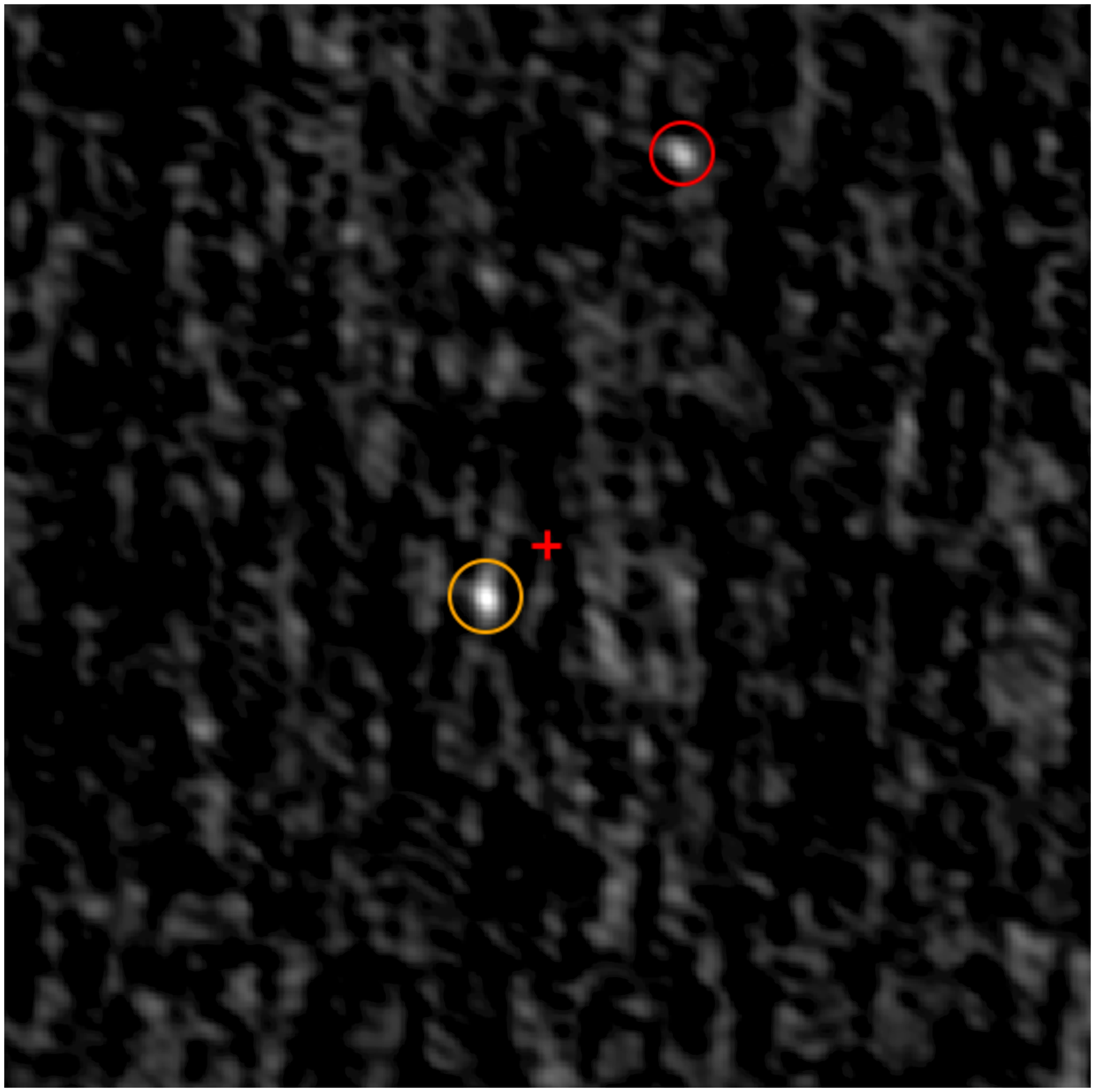}
\includegraphics[width=8cm]{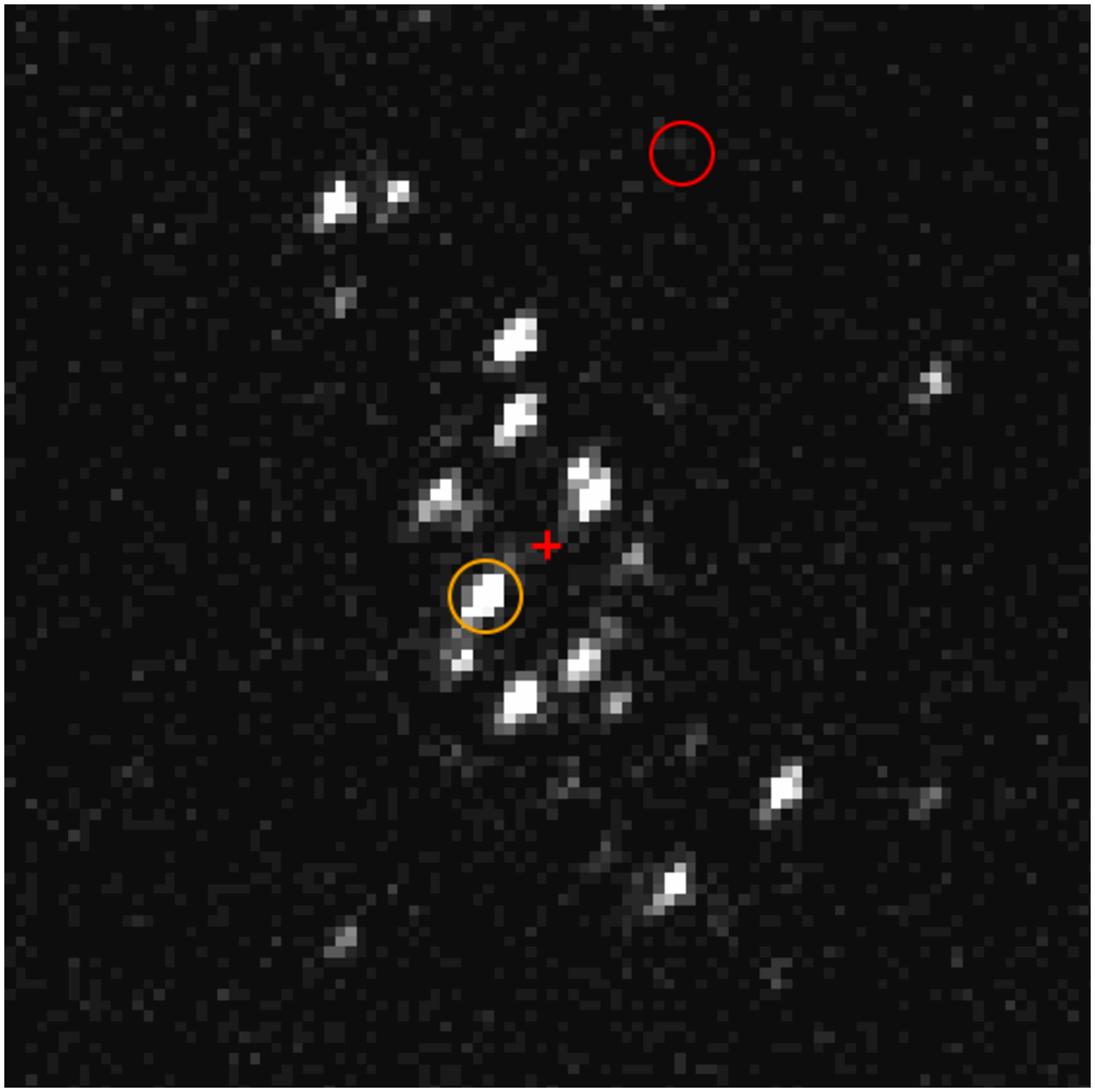}
\vspace{-1.1cm}
\caption{VLA 6.2 GHz radio (left) and \emph{Chandra} X-ray (right) images of the central 50\arcsec\ (1.6 pc) of M62, showing the candidate BH M62-VLA1 (orange circle). A red cross marks the cluster photometric center \citep{Lutzgendorf_etal13}. The other radio source in the central region of M62 is a known pulsar (red circle;  \citealt{Possenti_etal03}).
North is up and east is to the left.}
\label{m62}
\end{center}
\end{figure*}

Over the past $\sim$6 years, there have been solid BH candidates identified in GCs around other galaxies \citep[e.g.,][]{Maccarone_etal07, Irwin_etal10}. These sources are characterized as likely BH binaries because: (i) they have X-ray luminosities far above the Eddington luminosity for a single neutron star, and (ii) they vary significantly on short timescales, making it implausible that the luminosity originates from a superposition of several neutron star X-ray binaries. The most luminous of these objects is associated with a GC in the massive Virgo elliptical NGC 4472, and has a peak $L_X \sim 4\times10^{39}$ erg/s \citep{Maccarone_etal07}. Optical spectroscopy of the associated GC shows broad (1500 km s$^{-1}$) [\ion{O}{3}] emission but no Balmer lines, and the combined data is best explained by a model in which a stellar-mass BH is accreting at a super-Eddington rate from a CO white dwarf \citep{Zepf_etal08, Peacock_etal12}.

The few observed super-Eddington BHs in extragalactic GCs would then simply be those with the most extreme accretion rates. They likely represent only the very tip of the iceberg in terms of BH binaries in GCs. Many BHs with lower accretion rates are almost certain to exist among X-ray sources in GCs, but they are greatly outnumbered by neutron-star binaries and difficult to distinguish from neutron stars using X-ray data alone.

In \citet{Strader_etal12}, we developed a new strategy for identifying quiescent BH binaries in Milky Way GCs, making use of both radio and X-ray data (see also \citealt{Maccarone_Knigge07}). Stellar-mass BHs accreting at low rates have compact jets which emit radio continuum via partially self-absorbed synchrotron emission \citep{Blandford_Konigl79}. Thus they are much more luminous in the radio than neutron stars with comparable X-ray luminosity: $L_{R}/L_{X}$ is $\sim 2$ orders of magnitude higher for BHs than neutron stars \citep{Migliari_Fender06}. Before the recent upgrade to the VLA, the radio emission from a quiescent BH like A0620-00 or V404 Cyg would not have been detectable at high significance at typical GC distances \citep{Gallo_etal06}. The upgraded VLA can now readily detect the expected flux densities (tens of $\mu$Jy) in reasonable exposure times.

Using this technique, we discovered two candidate stellar-mass BHs in the cluster core of M22 \citep{Strader_etal12}. The sources have flat radio spectra and 6 GHz flux densities of 55--60 $\mu$Jy. As these sources are not detected in shallow archival \emph{Chandra} imaging, they cannot yet be placed directly on the $L_{X}-L_{R}$ relation; nevertheless, their overall properties are consistent with those expected from accreting BH binaries. 

Here we report the discovery of a BH candidate in a second Galactic GC, M62 (NGC 6266; $D = 6.8$ kpc; \citealt{Harris96}). We term this source M62-VLA1. Unlike the case for the M22 sources, M62-VLA1 has clear X-ray and optical counterparts, and so is the most compelling  candidate black hole X-ray binary in a Milky Way GC.

In Section 2, we describe our VLA observations, along with archival \emph{Chandra} and \emph{HST} imaging. In Section 3, we present evidence that M62-VLA1 is an accreting stellar-mass BH. Section 4 discusses the host binary system: the binary separation and binary companion. We assess alternative explanations for M62-VLA1 in Section 5. We summarize our findings in Section 6.

\section{Observations}
\subsection{VLA Radio Data}
We observed M62 with the Karl G.~Jansky Very Large Array (VLA) over the time period 2012 Sept 10--16 as part of the program 12B-073 (P.I.~Strader). Ten hours were spent observing the cluster, split among seven blocks of 1--1.75 hour duration, yielding a total of 7 hours on source. We observed in C band with 2 GHz total bandwidth and four polarization products. Of the two basebands of 1 GHz width, one was centered at 5.0 GHz and the other at 7.4 GHz. The array was in BnA configuration, giving a resolution of $1.4^{\prime\prime} \times 1.1^{\prime\prime}$ at 5.0 GHz. The field of view (full-width at half power) of the VLA at the average frequency of 6.2 GHz was $\sim$7.3$^{\prime}$ in diameter, significantly larger than the half-light diameter of M62 (1.8$^{\prime}$; \citealt{Harris96}).

We observed J1700-2610 as the secondary phase calibrator and J1407+2827 as the polarization leakage calibrator. 3C286 was used as an absolute flux density, bandpass, and polarization angle calibrator. The data were reduced using standard routines in AIPS. Weights from switched power measurements were applied using \verb|TYAPL|. Each observing block was edited for bad data and interference and then calibrated. For each individual calibrated baseband, the data were concatenated in the $uv$ plane and then imaged with a Briggs robust value of 1. A bright source at the edge of the field called for phase and amplitude self calibration. Figure \ref{m62} shows a deep co-added image of both basebands, obtained by smoothing the 7.4 GHz baseband to the resolution of the 5.0 GHz basebands and averaging these together. The rms sensitivity of this co-added image is 2.0 $\mu$Jy beam$^{-1}$.

M62-VLA1 is the only significant radio source in the core of M62. We use the $HST$-based core radius from \citet{Noyola_Gebhardt06} of $r_c = 6.6^{\prime\prime}$, and the updated center in J2000 coordinates of $17^{h}01^{m}13.0^{s}$, $-30^{\circ}06^{\prime}48.2^{\prime\prime}$ from \citet{Lutzgendorf_etal13}. M62-VLA1 is located at $17^{h}01^{m}13.217^{s}$, $-30^{\circ}06^{\prime}50.60^{\prime\prime}$, with positional uncertainty of 0.05$^{\prime\prime}$ in both coordinates. It is offset from the cluster center by 3.7$^{\prime\prime}$ (0.12 pc, assuming a distance to M62 of 6.8 kpc; \citealt{Harris96}).  There is an additional radio source somewhat outside M62's core, at a cluster radius of 19.1$^{\prime\prime}$. It is a known binary millisecond pulsar, PSR J1701-3006A \citep{Possenti_etal03}.  

Flux density measurements were carried out separately on each baseband's image. We fit Gaussians using \verb|JMFIT| in AIPS assuming the source is point-like.  Measured flux densities for M62-VLA 1 are $19.9\pm3.2$ $\mu$Jy (5.0 GHz) and $18.1\pm2.3$ $\mu$Jy (7.4 GHz). Assuming the flux densities follow a power law of the form $S_{\nu} = \nu^{\alpha}$, then the spectral index of M62-VLA1 is $\alpha=-0.24\pm0.42$. In contrast, PSR J1701-3006A was measured to have a 5.0 GHz flux density of $23.6\pm2.9$ $\mu$Jy, but was not detected at 7.4 GHz ($3.7\pm2.3$ $\mu$Jy), implying $\alpha \lesssim -2.0$. The spectrum of M62-VLA1 is inconsistent with the steep spectrum expected for a pulsar.

\begin{deluxetable*}{cccccccccccccc}
\tablewidth{0 pt}
\tabletypesize{\footnotesize}
\setlength{\tabcolsep}{0.025in}
\tablecaption{ \label{xfit}
Model Fits M62-VLA1 in \emph{Chandra} Data }
\tablehead{Model & $N_{H}$\tablenotemark{a} & $\Gamma$ & $L_{X}$\tablenotemark{b} & kT & $L_{X, {\rm therm}}$ & $\chi^{2}/\nu$ \\
& ($10^{21}$ cm$^{-2}$) & & (erg s$^{-1}$) & (keV) &  (erg s$^{-1}$) & }
\startdata
power-law & 2.8 & $2.5\pm0.1$ & $3.3\times 10^{32}$ & --- & --- & 37.7/29 \\
power-law + blackbody & 2.8 & $1.9\pm0.2$ & $5.9\times 10^{32}$ & $0.17\pm0.02$ & $1.8\times 10^{32}$ & 28.1/27 \\
power-law + NS atmosphere & 2.8 & $1.8\pm0.3$ & $5.9\times 10^{32}$ & $0.08\pm0.01$ & $2.1\times 10^{32}$ & 27.5/27 
\enddata
\tablenotetext{a}{Held fixed at the estimated foreground value. $^{\rm b}$ In the energy range 0.5--7 keV.}
\end{deluxetable*}

\subsection{\emph{Chandra} X-ray Data} \label{xrayobs}
The archival \emph{Chandra} imaging was obtained in May 2002 with ACIS-S and a total integration time of 63 ksec (Obs ID 2677, P.I.~Lewin). M62 hosts a rich population of X-ray sources, with 51 sources detected within the half-mass cluster radius in the \emph{Chandra} image and only 2--3 expected to belong to the background \citep{Pooley_etal03}. However, the properties of the individual sources have not been published.

M62-VLA1 is the only X-ray source in M62 which has a counterpart in our VLA image (Figure \ref{m62}). It is detected in the \emph{Chandra} image at high significance, with 688 counts or signal-to-noise ratio of 26.

We fit the X-ray spectrum of M62-VLA in the energy range 0.5--7 keV (experiments with other energy ranges yielded very similar results). The foreground absorption was held constant at $N_{H} = 2.8 \times 10^{21}$ cm$^{-2}$, calculated from the reddening estimate for M62 ($E(B-V) = 0.47$; \citealt{Harris96}) and using the \citet{Guver_etal09} relation between reddening and $N_{H}$. The spectrum was binned so that there were at least 20 counts per channel, and then fit via $\chi^2$ minimization in XSPEC.

Assuming a power-law form to the spectrum (as expected for an accreting BH in the low/hard state), we found a photon index $\Gamma = 2.5\pm0.1$ (90\% confidence interval) and a 0.5--7 keV luminosity of $\sim$3$\times 10^{32}$ erg s$^{-1}$ (Table \ref{xfit}).  The fit has a 13\% null hypothesis probability ($\chi^2/{\nu} = 1.30$ with 29 degrees of freedom).

The spectrum was not well-fit by an absorbed blackbody ($\chi^2/{\nu} = 5.17$ with 29 degrees of freedom). However, we did find a reasonable fit with a combined blackbody and power-law spectrum, as observed for accreting neutron stars. The parameters for this model are listed in Table \ref{xfit}; the power law component accounts for about $\sim$2/3 of the flux and the best-fit blackbody temperature is $1.9 \times 10^{6}$ K. A similarly good fit can be achieved by fitting a model neutron star atmosphere plus a power law, using a NSATMOS model in XSPEC \citep{Heinke_etal06} and assuming a neutron star mass of 1.4 M$_{\odot}$ and radius of 12 km. The best-fit parameters are again listed in Table \ref{xfit}; the luminosity is practically identical to the power-law + blackbody model, but the temperature of the best-fit neutron star atmosphere is lower.

These fits emphasize the well-known conclusion that it is challenging to distinguish BHs from neutron stars on the basis of X-ray spectra alone.

\begin{figure}[t]
\begin{center}
\vspace{-0cm}
\includegraphics[angle=90, width=3in]{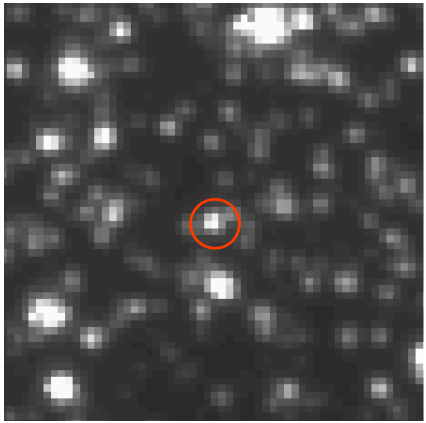}
\vspace{0cm}
\caption{$HST$ $F435W$ image of the candidate optical counterpart to M62-VLA1. The astrometric matching radius (0.17\arcsec) is shown as a red circle. North is up and east is to the left; the image is about 3\arcsec\ on a side.}
\label{opt}
\end{center}
\end{figure}

\subsection{HST Optical Data}

There are archival \emph{Hubble Space Telescope (HST)}/Advanced Camera for Surveys (ACS) data that cover the core of M62, obtained in 2004 (two years after the \emph{Chandra} data discussed in \S 2.2, but more than eight years before the VLA data were observed). These data comprised 880 sec of exposure time in $F435W$ ($B$-equivalent; split into three frames), 1170 sec in $F625W$ ($R$-equivalent; five frames), and 3610 sec in $F658N$ (narrow 
band covering H$\alpha$; ten frames). 

We photometered the individual frames using DOLPHOT \citep{Dolphin00} before producing a matched photometric catalog. We note that M62 has a large and differential foreground reddening with mean $E(B-V)=0.47$ \citep{Alonso-Garcia_etal11}; all photometry quoted in this paper is uncorrected for reddening.

To match the astrometry of the VLA images, we iteratively corrected the astrometry of the $HST$ images to the ICRS using a large number of 2MASS sources across the images. The overall uncertainty in the astrometric zero point of the image is about 0.15\arcsec. The uncertainty in the position of M62-VLA1 itself is 0.07\arcsec, giving a combined $1\sigma$ uncertainty of 0.17\arcsec\ in matching the $HST$ and VLA data. Unfortunately there are no additional sources that appear in both the $HST$ and VLA or \emph{Chandra} images, so further checks on the astrometry are not possible.

\begin{figure}
\begin{center}
\vspace{-1.1cm}
\includegraphics[width=3.2in]{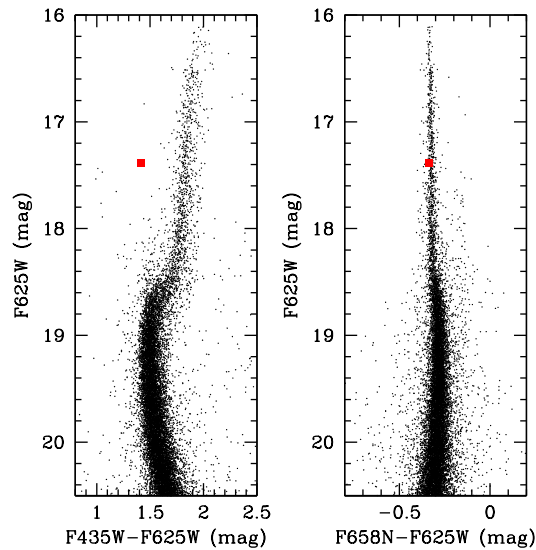}
\vspace{-2cm}
\caption{$HST$ color-magnitude diagrams in $F625W$ vs.~$F435W-F625W$ (left panel) and $F625W$ vs.~$F658N-F625W$ (right panel) of stars in M62. The mean photometry of the M62-VLA1 candidate optical counterpart discussed in \S 2.3 is plotted as a red square. The horizontal branch would be located at $F625W \la 16$ mag, but most of these stars are saturated in the $F625W$ photometry.}
\label{opt2}
\end{center}
\end{figure}

Figure \ref{opt} shows the position of M62-VLA1 superposed on the $HST$ $F435W$ image. Even given the significant astrometric uncertainty, M62-VLA1 is a very close ($< 0.05$\arcsec) match to a moderately bright star in the images. This object has mean $F625W = 17.38$ mag, and $F435W-F625W = 1.41$. In a $F625W$ vs.~$F435W-F625W$ color-magnitude diagram the star falls near the lower red giant branch of M62, but lies $\sim$0.4 mag blueward of the cluster fiducial (Figure \ref{opt2}). 

\begin{figure}
\begin{center}
\vspace{-1.35cm}
\includegraphics[width=3.5in]{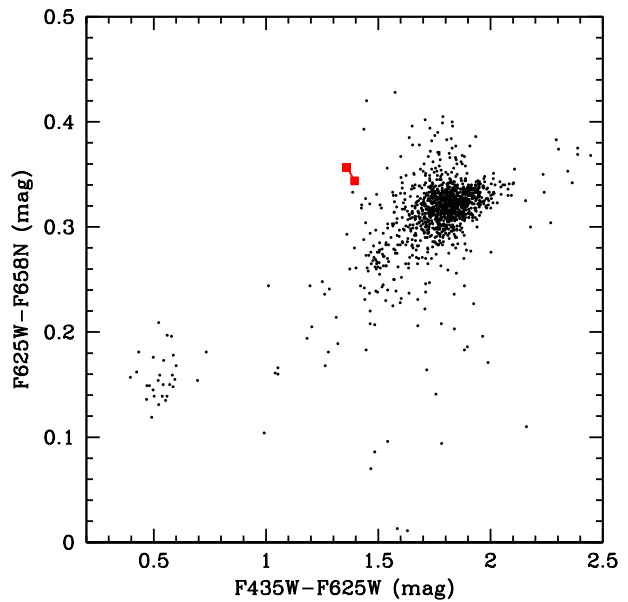}
\vspace{-2.5cm}
\caption{\emph{HST} $F625W-F658N$ vs.~$F435W-F625W$ color-color diagram of unsaturated stars with $F625W < 19$ mag in the central regions of M62. The candidate optical counterpart of M62-VLA1 (red squares; data at two epochs) is $\sim$0.09 mag more luminous in $F658N$ than the main locus of stars. One explanation is that this deviation is due to the presence of H$\alpha$ emission in the $F658N$ band with an equivalent width of $\sim$7 \AA.}
\label{colcol}
\end{center}
\end{figure}

The $F625W-F658N$ color of the star is $\sim$0.08 to $0.1$ mag redder than the inferred locus of $F435W-F625W$ giants, calculated by connecting the principal population of red giant branch stars to that of horizontal branch stars. This photometry is consistent with the presence of H$\alpha$ emission with an equivalent width of $\sim$7 \AA\ in the $F658N$ filter. Figures \ref{opt2}  and \ref{colcol} show color-magnitude and color-color diagrams with the position of this star marked.

This brighter star is flanked by two fainter stars that also fall just within the nominal astrometric error circle (red region in Figure \ref{opt}). These other stars have photometry consistent with main sequence stars just below the turnoff of M62.

\begin{figure}
\begin{center}
\vspace{-0.35cm}
\includegraphics[width=3.5in]{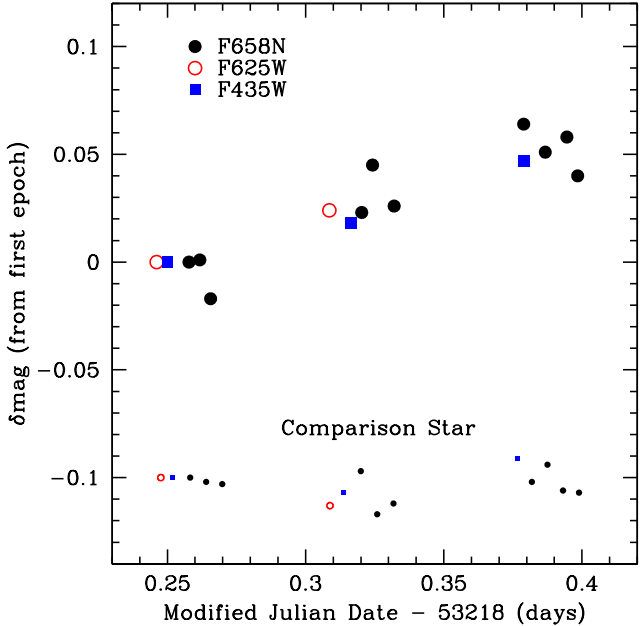}
\caption{\emph{HST} optical variability of the likely counterpart to M62-VLA1. The ACS data, taken over 0.15 days (3.6 hr), shows a clear linear trend in three different filters. The star is more variable than 98\% of stars of comparable brightness, lending credence to the interpretation that the source is an accreting binary. Photometry of a randomly chosen comparison star of similar $F658N$ magnitude is shown at bottom with an arbitrary offset.}
\label{varz}
\end{center}
\end{figure}

There is some evidence that the brighter candidate optical counterpart star is variable. The ten frames of $F658N$ photometry, acquired over 3 hours, show a linear trend to fainter magnitudes, with amplitude $\sim$0.06 mag over this time period (see Table \ref{hstphot}, which lists all photometry of this star, and Figure \ref{varz}). To assess the significance of this trend, we examined the photometry of stars of comparable brightness ($16.5 < F658N < 17.5$ mag). For these comparison stars, on average, there is no evidence for a systematic trend in the photometry among the frames. If we fit linear relations to the $F658N$ magnitudes as a function of time for all the comparison stars, the inferred amplitude of the trend for the M62-VLA1 candidate counterpart is larger than 98\% of the stars. The three $F435W$  frames cover a comparable time baseline and show a similar trend, although with a somewhat smaller amplitude. Due to saturation, only two of five frames in $F625W$ have reliable photometry, but these are consistent with the behavior observed in the other filters.

Because of the crowding in the core of M62 and the relatively modest amplitude of the variability, we consider this finding worthy of further investigation but far from conclusive.

\begin{deluxetable}{ccc}
\tablewidth{0 pt}
\tabletypesize{\footnotesize}
\setlength{\tabcolsep}{0.025in}
\tablecaption{ \label{hstphot}
Photometry of Candidate Optical Counterpart to M62-VLA1}
\tablehead{MJD\tablenotemark{a} & Phot. & Filter \\
(days) & (mag) & }
\startdata
53218.25824 & 17.021 & $F658N$ \\
53218.26403 & 17.022 & $F658N$ \\
53218.26983 & 17.004 & $F658N$ \\
53218.31998 & 17.044 & $F658N$ \\
53218.32590 & 17.066 & $F658N$ \\
53218.33181 & 17.047 & $F658N$ \\
53218.38188 & 17.085 & $F658N$ \\
53218.38750 & 17.072 & $F658N$ \\
53218.39313 & 17.079 & $F658N$ \\
53218.39892 & 17.061 & $F658N$ \\
\hline
53218.25199 & 18.773 & $F435W$ \\
53218.31368 & 18.791 & $F435W$ \\
53218.37653 & 18.820 & $F435W$ \\
\hline
53218.24763 & 17.372 & $F625W$ \\
53218.30880 & 17.396 & $F625W$ 
\enddata
\tablenotetext{a}{Modified Julian Date at midpoint of exposure.}
\tablecomments{All photometry is on the VEGAMAG system and is not corrected for foreground reddening. Uncertainties on the photometry (dominated by systematic
effects in the crowded core) are $\sim$0.01 mag.}
\end{deluxetable} 

\begin{figure*}
\begin{center}
\includegraphics[angle=90, width=6in]{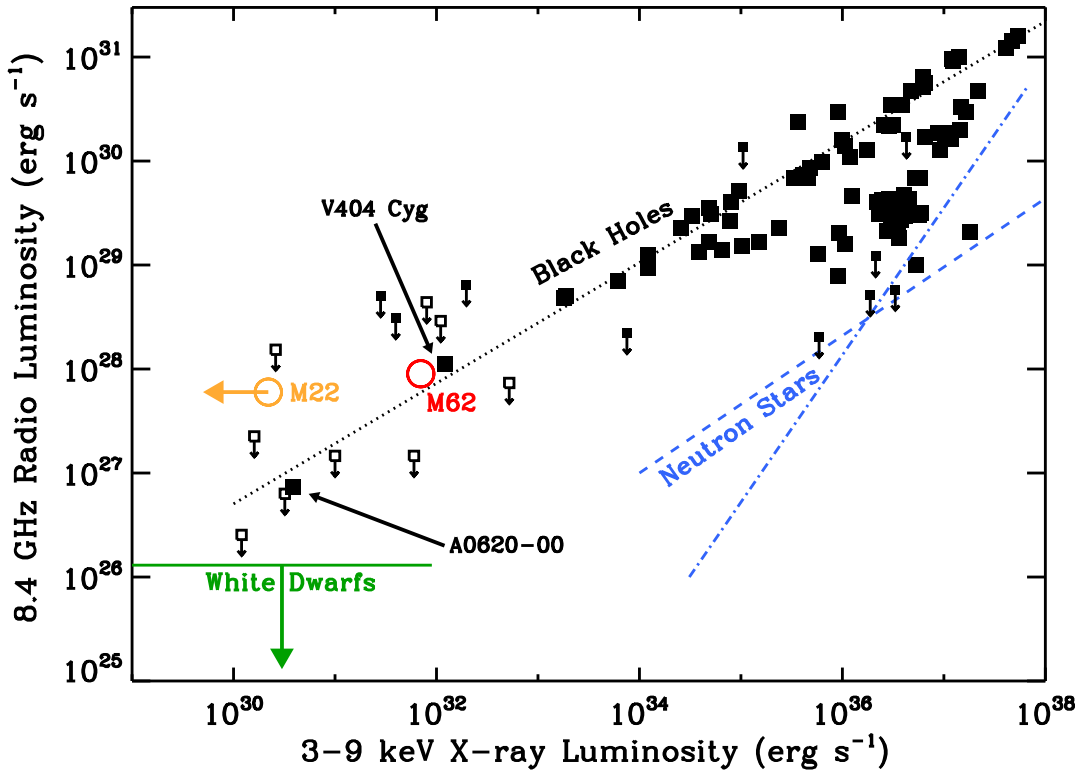}
\caption{The radio--X-ray correlation for stellar-mass BHs, showing M62-VLA1 as an open red circle. The open orange circle represents our two BH candidates in M22, which have very similar radio luminosities to one another and have not yet been detected in X-rays \citep{Strader_etal12}. Filled squares have simultaneous X-ray and radio observations; unfilled points are non-simultaneous. Black points are stellar-mass BHs from the literature \citep{Miller-Jones_etal11, Gallo_etal12, Ratti_etal12, Corbel_etal13}; some BHs have multiple measurements plotted, representing different accretion phases. The dotted black line is the BH correlation of \citet{Gallo_etal06} ($L_R \propto L_X^{0.58}$, normalized by a least-squares fit to the simultaneous detections with $L_X < 2\times10^{34}$ erg s$^{-1}$). The blue lines are two possible correlations for accreting neutron stars \citep{Migliari_Fender06}. The solid green line is the maximum radio continuum luminosity detected for accreting white dwarfs \citep{Kording_etal08, Kording_etal11}.}
\label{lxr}
\end{center}
\end{figure*}

\section{A New Black Hole Candidate}

The most convincing evidence supporting a BH identification for M62-VLA1 is its ratio of X-ray luminosity ($L_{X}$) to radio luminosity ($L_{R}$). Figure \ref{lxr} plots M62-VLA1 on the standard X-ray--radio correlation \citep{Gallo_etal06}. Its mean flux density of 19 $\mu$Jy corresponds to an equivalent 8.4 GHz radio luminosity of $L_{R} = 9 \times 10^{27}$ erg s$^{-1}$ at the distance of M62, assuming a flat spectrum ($\alpha = 0$; if the measured $\alpha=-0.24$ is used instead, then $L_{R} = 8 \times 10^{27}$ erg s$^{-1}$). A spectral fit to the X-ray data gives a luminosity of $L_{X} = 7 \times10^{31}$ erg s$^{-1}$ over 3--9 keV. This particular radio frequency and X-ray spectral range are chosen because they are the most commonly used in $L_{X}-L_{R}$ relations in the literature.

With the important caveat that variability could be present between the X-ray and radio epochs, M62-VLA1 sits squarely on the BH  $L_{X}-L_{R}$ relation. The source appears to be a doppelganger for the quiescent BH V404~Cyg in its X-ray and radio luminosities \citep{Miller-Jones_etal09}. The radio luminosity for M62-VLA1 is $\ga 2$ orders of magnitude higher than expected for accreting neutron stars or white dwarfs.

Both the radio and X-ray spectra of M62-VLA1 are consistent with an accreting stellar-mass BH. The radio spectral index is consistent with being flat ($\alpha=-0.2\pm0.4$), similar to the radio indices of known low-luminosity accreting stellar-mass BHs \citep[$\alpha$ = 0.0--0.2;][]{Fender01, Gallo_etal05}. However, the radio spectral index could also be consistent with a flaring neutron star, which can have a radio spectrum that varies from  $\alpha \approx -0.6$ as expected for optically-thin synchrotron emission to slightly inverted ($\alpha > 0$; e.g., \citealt{Marti_etal92, Fender97, Moore_etal00, Miller-Jones_etal10}).

As described in Section \ref{xrayobs}, we also cannot rule out a neutron star origin for M62-VLA1 from the X-ray spectra alone. Either a single power-law or  a power-law + thermal model is a good match to the X-ray data. The power-law spectral fit for M62-VLA1 ($\Gamma = 2.50\pm0.14$) is comparable to that of the quiescent BH binaries V404 Cyg ($\Gamma = 2.17\pm0.13$; \citealt{Corbel_etal08}), A0620-00 ($\Gamma = 2.1^{+0.5}_{-0.4}$; \citealt{Gallo_etal06}), and MAXI J1659--152 ($\Gamma = 2.5\pm0.3$; \citealt{Jonker_etal12}), although it is slightly softer than commonly observed. The X-ray spectrum of M62-VLA1 is consistent with the claim that the low-luminosity BHs display a softening of their X-ray spectra in quiescence \citep{Corbel_etal06,Wu_Gu08}. 

At a projected distance of 3.7$^{\prime\prime}$\ from the photometric center (0.12 pc at the distance of M62), the separation of M62-VLA1 is too large to
be consistent with an intermediate-mass BH ($\sim$10$^{2}-10^{4}$ M$_{\odot}$): such sources would be expected to be $\la 1\arcsec$ from the center \citep{Chatterjee_etal02, vanderMarel_Anderson10, Strader_etal12a}. A caveat is that the photometric center could be offset from the dynamical center, but other evidence disfavors an intermediate-mass BH interpretation. For example, the fundamental plane of accreting BHs \citep{Merloni_etal03} implies that an intermediate-mass BH should have a much higher $L_{R}/L_{X}$ than a stellar-mass BH.

\section{The Host Binary System}

Assuming that M62-VLA1 is a stellar-mass BH in M62, the intracluster gas density is very unlikely to be high enough to produce the observed radio and X-ray luminosity via Bondi accretion. Therefore, M62-VLA1 must be accreting from a binary companion. Assuming a nominal BH mass of 10 M$_{\odot}$, M62-VLA1 would be accreting at $\sim (2-3) \times 10^{-7}$ L$_{\rm edd}$.

The relatively high quiescent X-ray luminosity of M62-VLA1, coupled with the period--luminosity correlation for BH binaries \citep{Garcia_etal01, Reynolds_Miller11}, suggests a longer orbital period than expected for a dwarf secondary. A giant donor is more likely. This would be yet another similarity between M62-VLA1 and V404~Cyg (which has a long orbital period for a BH binary at 6.5 days; \citealt{Casares_etal92}). 

Consistent with this interpretation, the $HST$ data discussed in \S 2.3 reveals a candidate optical counterpart consistent with a star on the lower red giant branch of M62. However, the $F435W-F625W$ color of the star is unusually blue for an M62 giant, which could indicate the contribution of thermal emission from an accretion disk around the black hole. The photometry also suggests the presence of H$\alpha$ emission, which could arise from an accretion disk. 

In fact, the properties of the possible optical counterpart to M62-VLA1---including its unusual color---present a reasonable resemblance to V404~Cyg. \citet{Casares_etal93} estimate that for V404~Cyg, $\sim$26\% and $\sim$9\% of the $B$ and $R$-band light, respectively, is attributable to accretion. Given these values, which are uncertain by perhaps a factor of two, the star is more than 0.2 mag bluer in $B-R$ than it would be without the presence of an accretion disk (compare this to the $\sim$0.4 mag offset inferred for M62-VLA1). The equivalent width of H$\alpha$ emission in V404~Cyg is $\sim$20 \AA, although somewhat time variable \citep{Hynes_etal02, Hynes_etal09}. From our broadband and $F658N$ photometry of M62-VLA1, we estimate a rough H$\alpha$ equivalent width of $\sim$7 \AA.

Optical variability often accompanies accreting binaries (see \citealt{Edmonds_etal03} for binaries in 47 Tucanae), and the $HST$ variability observed in all three filters (including the $F658N$ filter covering H$\alpha$; see \S 2.3 and Figure \ref{varz}) supports the identification of the candidate giant counterpart with the radio source M62-VLA1.

For comparison, consider V404~Cyg, which exhibits variability on a wide range of time scales, from minutes to days \citep{Casares_etal93, Pavlenko_etal96, Shahbaz_etal03, Zurita_etal04}. In addition to its 6.5-day orbital period that produces ellipsoidal variations, it shows a $\sim$6-hr quasi-periodic oscillation with variable amplitude, 0.05--0.25 mag \citep{Pavlenko_etal96}; a similar mode might explain the variability plotted in Figure \ref{varz}. In addition, when V404~Cyg varies the H$\alpha$ and optical continuum usually scale together \citep{Hynes_etal02}; the behavior is also seen in the proposed optical counterpart to M62-VLA1 (Figure \ref{varz}).

The association between M62-VLA1 and this optical source could be bolstered by further high-resolution photometry, follow-up spectroscopy, and a proper-motion measurement. It should be kept in mind that M62-VLA1 could instead be associated with a star not detected in the $HST$ images, such as an M dwarf---although this scenario might be disfavored, as the shorter orbital period would lead to a lower mass accretion rate and thus a lower quiescent X-ray luminosity than observed. The system could also, in principle, be a white dwarf-BH binary at an orbital period such that its quiescent luminosity would match the observations; the first extragalactic BH X-ray binary candidate in a GC is likely to be such a system \citep{Maccarone_etal07, Zepf_etal08}. 

\section{Alternative Explanations}

While the properties of M62-VLA1 are well-described by an accreting stellar-mass BH in M62, here we consider other possible scenarios which might produce this source. Additional discussion for some of these possibilities can be found in the Supplementary Information of \citet{Strader_etal12}.

\subsection{Accreting Neutron Star or White Dwarf}

As shown in Figure \ref{lxr}, neutron stars or white dwarfs accreting from binary companions are expected to have much lower $L_R/L_X$ and $L_R$ than M62-VLA1 \citep{Fuerst_etal86, Migliari_Fender06, Kording_etal08, Kording_etal11, Byckling_etal10}. An exception is the class of symbiotic stars, which are white dwarfs accreting from luminous red giants. Radio emission in symbiotic stars is usually thermal, originating from the ionized red giant wind. Therefore, high $L_R/L_X$ values are observed in these systems (log~$L_{R}/L_X \approx -2$ to $-1$, defined for a radio frequency of 6.2 GHz and X-ray energy range 0.5--8 keV; \citealt{Maccarone_etal12}). This ratio is significantly higher than observed for M62-VLA1 (log~$L_{R}/L_X = -4.7$). Symbiotic stars also show inverted radio spectra ($\alpha \approx 0.6$; \citealt{Seaquist_Taylor90}), inconsistent with M62-VLA1. Finally, the giant star which has been tentatively associated with M62-VLA1 is also less optically luminous than typical symbiotic stars. For these reasons, combined with the fact that there are no known symbiotic stars in GCs, we consider it unlikely that M62-VLA1 is a symbiotic star.

We can offer a contrived scenario that could make the observations of M62-VLA1 consistent with that for an accreting neutron star, if we assume the source is strongly variable. The radio and X-ray data on M62 were taken 10 years apart. The ``high" radio luminosity could represent a strong flare on a neutron star, while the accompanying rise in $L_X$ may have remained below the detection threshold of X-ray all-sky monitors. The \emph{Chandra} data would then need to represent a quiescent phase. We note that the duty cycle of neutron star flares is probably low but is poorly constrained.

\begin{figure}
\begin{center}
\vspace{-0.5cm}
\includegraphics[angle=90, width=3.4in]{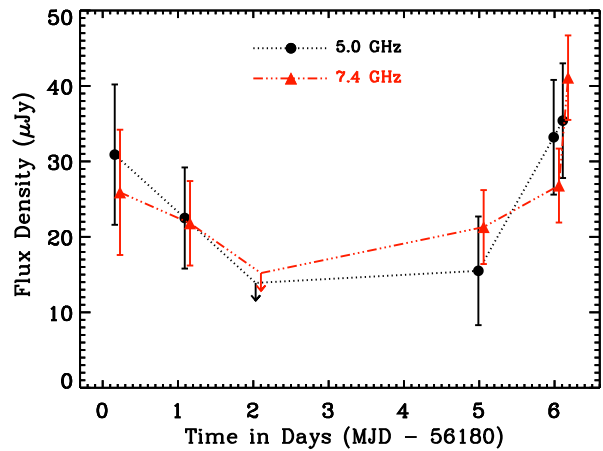}
\caption{Radio light curve of M62-VLA1 over the week of VLA observations. Black circles connected by a dotted line are 5.0 GHz measurements, while red triangles linked with a dot-dashed line are 7.4 GHz data. Observations are simultaneous at the two frequencies, but the 7.4 GHz data points are offset in time by 0.07 days for clarity. The third epoch, observed on MJD 56182.0, is a non-detection at both frequencies. }
\label{radlc}
\end{center}
\end{figure}

Strong neutron star outbursts are usually accompanied by a rapid rise and decline in radio luminosity on short timescales (factor of $\ga 3$ in $\sim$4 days; \citealt{Moore_etal00, Tudose_etal09, Miller-Jones_etal10}). To test this predicted evolution, we imaged each of the VLA observing blocks separately and measured the flux density of M62-VLA1 in six distinct epochs spanning a week (the second and third observing blocks were scheduled directly adjacent to one another and combined into one epoch); the light curve is plotted in Figure \ref{radlc}. M62-VLA1 varies by a factor of $\ga 3$, although its variations are not in the form of a single flare. Instead, the flux density drops over the first two days to a non-detection during the third epoch, and then rises again to reach a maximum in the last epoch. The light curve of M62-VLA1 therefore appears to be inconsistent with a single strong outburst of a neutron star, and is more consistent with the ``burbling" observed for quiescent stellar-mass black holes \citep{Miller-Jones_etal08}. Still, for a secure interpretation of M62-VLA1, simultaneous X-ray and radio data are highly desirable, and are being pursued.

\subsection{Pulsar or Supernova Remnant}
Millisecond pulsars are often found in the cores of GCs, but they display steep radio spectra ($\alpha \lesssim -1$; \citealt{Kramer_etal99}), inconsistent with the measured spectral index of M62-VLA1.

Pulsars in the field occasionally drive a wind that interacts with ambient material, termed a pulsar wind nebula. These objects have flat radio spectra, but again are more radio loud than M62-VLA1 (log~$L_{R}/L_X \approx 0$; \citealt{Maccarone_etal12}). In addition, they generally have $L_X > 10^{34}$ erg s$^{-1}$, large diameters, short lifetimes, and dense surroundings, all inconsistent with observations of M62-VLA1 \citep{Gaensler_Slane06}. A similar analysis would apply to ``redback"  systems in which the material originates in an ablated companion \citep{Roberts13}.

Like pulsar wind nebulae, supernova remnants should be short lived, angularly extended, and located in dense environments. Supernova remnants in our Milky Way are more radio luminous than M62-VLA1 ($>$0.1 Jy) and have steeper radio spectra ($\alpha \approx -0.7$; \citealt{Green09}). For these reasons, we exclude pulsars, pulsar wind nebulae, and supernova remnants as likely identities of M62-VLA1.

\subsection{Planetary Nebula}
Planetary nebulae emit optically-thin thermal emission at radio wavelengths, and would therefore show a flat spectral index consistent with M62-VLA1. However, planetary nebulae have low X-ray luminosities ($\sim$10$^{30}$~erg~s$^{-1}$; \citealt{Montez_etal10}), much fainter than M62-VLA1. In addition, many planetary nebulae are bright in H$\alpha$, but no nebulosity is observed around M62-VLA in archival $HST$/WFPC2 images obtained in the F656N filter. 

\subsection{Foreground Active Star}
Foreground ($D \lesssim 100$ pc) coronally-active late-type stars can be bright in both the X-ray and radio regimes, show flat radio spectra, and have log~$L_{R}/L_X \approx -6$ to $-2$ \citep{Berger_etal10, Maccarone_etal12}. The candidate optical counterpart of M62-VLA1 is a K giant, not a very late-type dwarf. On the other hand, this match could be spurious and an active dwarf star might escape undetected in the optical imaging if sufficiently distant ($\gtrsim$ 50 pc). To produce detectable radio and X-ray emission at these distances, the star would need to be flaring \citep{Berger_etal10}. However, a flaring star should also be strongly circularly polarized \citep{Berger02, Hallinan_etal08}, and there is no evidence that M62-VLA1 is circularly polarized ($3\sigma$ upper limit 22\%). 

\subsection{Background Galaxy}
It is possible, but very unlikely, that M62-VLA1 is a background radio galaxy. Galaxies exhibit a wide range of log~$L_{R}/L_X$ from $0$ to $-5$ \citep{Maccarone_etal12}. In addition, they show a diversity of radio spectral slopes, although steeper indices than measured for M62-VLA1 are typical ($\alpha \approx -0.7$; \citealt{Fomalont_etal02, Randall_etal12}). Background galaxies with flat radio spectra are either active galactic nuclei with partially self-absorbed jets, or star-forming galaxies in which the combination of synchrotron and thermal emission produces a flat composite spectrum. Background objects with flux densities as faint as M62-VLA1 ($\sim$20 $\mu$Jy) have not yet been characterized in detail.

We use calculations similar to those in \citet{Strader_etal12} to determine the expected number of background sources with properties comparable to those of M62-VLA1. \citet{Fomalont_etal02} estimate that there are 0.3 sources per sq. arcmin with 8.4 GHz flux densities in the range 10--30 $\mu$Jy. We would therefore expect $\sim$0.01 sources in the 6.6$^{\prime\prime}$ radius core of M62 with similar flux densities as M62-VLA1. This number decreases to $\sim$0.003 when the spectral index is considered; \citet{Fomalont_etal02} find that only 30\% of the 47 faint radio sources ($< 35\ \mu$Jy) in their sample have $\alpha \geq -0.4$, measured from VLA 1.4 and 8.4 GHz data. The lack of a visible background galaxy further reduces the probability that M62-VLA1 is a background object: if we conservatively assume that we could detect optical counterparts to $I \approx 21.5$, about a third of the faint flat-spectrum radio sources in the Fomalont et al.~sample would have detectable optical matches. On the basis of the radio and optical data, we therefore estimate that the probability that M62-VLA1 is a background galaxy is $< 0.003$. 

The X-ray counterpart to M62-VLA1 further decreases the probability that it is a background galaxy. For $2-10$ keV fluxes equivalent to or brighter than M62-VLA1 ($S_{\rm 2-10 keV} > 2 \times 10^{-14}$ erg s$^{-1}$ cm$^{-2}$; \citealt{Mateos_etal08}), only $\sim$10$^{-3}$ X-ray sources are expected in the core of M62. The X-ray spectrum ($\Gamma = 2.5\pm0.1$) is also unusually soft for background sources. Of relatively bright background sources ($S_{\rm 2-10 keV} \ga 3 \times 10^{-15}$ erg s$^{-1}$ cm$^{-2}$), only $\sim$10\% are fit with $\Gamma \ga 2.2$ \citep{Mateos_etal05, Page_etal06, Lanzuisi_etal13}. Therefore, we estimate a very low probability, $\sim$10$^{-4}$, that the X-ray counterpart to M62-VLA1 is associated with a background galaxy.

To date we have only investigated four GCs with deep VLA imaging (\citealt{Strader_etal12a}, plus M62); two of these GCs have candidate stellar-mass BHs. Therefore it is unlikely that the detection of a BH look-alike in M62 is due to a ``multiple trials" effect. Nonetheless, a direct elimination of the background-galaxy possibility is still needed. As in the accreting neutron star scenario, simultaneous radio and X-ray observations will help distinguish between a stellar-mass BH and a background galaxy.  The fundamental plane of BH activity suggests that $L_{R}/L_X \propto M_{\rm BH}^{0.38}$ \citep{Merloni_etal03}. If simultaneous data show a similar ratio as currently observed, then M62-VLA1 is unlikely to be a supermassive BH in a background active galactic nucleus. In addition, deeper $HST$ photometry would allow us to search for faint background galaxies coincident with M62-VLA1. Finally, deep high-resolution radio imaging with the High Sensitivity Array would allow us to test if M62-VLA1 shares the proper motion of M62 itself \citep{Dinescu_etal03}.

\section{Conclusions and Implications}

M62-VLA1 is the only radio source in the core of M62 down to flux density levels of $\sim$6 $\mu$Jy at 5--10 GHz. We have argued that it is likely to be a stellar-mass BH accreting from a binary companion. This scenario is supported by its X-ray and radio luminosities and spectra. The secondary may be a giant star near the base of the red giant branch, as suggested by $HST$ imaging and the $L_{X}$--period relation for known BHs.

While all of our data are consistent with a stellar-mass BH interpretation, other explanations are possible (although unlikely). The most plausible alternative scenarios are (i) an accreting neutron star X-ray binary that is in quiescence during the \emph{Chandra} observations but flaring during the VLA observations; or (ii) a background galaxy. Future observations---principally simultaneous X-ray and radio imaging---will conclusively distinguish between these scenarios.

M62 is now the second Milky Way GC with significant evidence for stellar-mass BHs (see \citealt{Strader_etal12} for discussion of the BH candidates in M22). We have identified BH candidates in two out of four GCs with deep VLA radio imaging. While the sample is certainly small, these findings suggest that stellar-mass BHs may be present in many GCs. However, the current sample is not representative of the mass or core density distribution of the Galactic GC system, so our findings are not yet easily generalized. We note that both M62 and M22 are massive GCs, but their structural parameters are quite different. While M22 has a large core radius, M62 has a relatively dense, compact core. This contrast may imply that there are no clear structural diagnostics of the presence BHs in GCs (in contrast with theoretical findings by e.g., \citealt{Mackey_etal08, Sippel_Hurley13, Morscher_etal13}).

It is worth emphasizing that the BHs which can be detected in accreting binary systems are likely to represent only a small fraction of the population of BHs in a star cluster, since single BHs and those in binaries not undergoing mass transfer may dominate the BH population in a GC (e.g., \citealt{Ivanova_etal10}).

If indeed many GCs host substantial populations of BHs, this represents an important shift in our view of stellar remnants in GCs. The potential implications are large: 
GCs would become good hunting grounds for new stellar-mass BHs, of which few are still confirmed; 
BHs in GCs are likely to be more massive than those in the field, opening up a new area of BH parameter space; 
it would be possible to determine accurate physical parameters for a significant sample of BHs, since the distances to GCs are accurately known; 
additional quiescent BH systems would offer new opportunities to investigate the physics of accretion at the lowest luminosities; 
and if GCs have multiple BHs, the likelihood of the dynamical formation of BH--BH and BH--pulsar systems will increase, improving prospects for the detection of gravitational waves and stringent tests of general relativity. 
Deep radio imaging of additional GCs, combined with detailed follow-up observations of the candidate BHs in M62 and M22, will determine whether GCs---long thought to be among the least likely hosts for BHs---turn out to be among the best.

\acknowledgements
We thank S.~Sigurdsson and N.~Ivanova for useful comments. L.~Chomiuk is a Jansky Fellow of the National Radio Astronomy Observatory. J.~C.~A.~Miller-Jones acknowledges support from an Australian Research Council Discovery Grant (DP120102393). C.~Heinke is supported by NSERC and an Ingenuity New Faculty Award. The National Radio Astronomy Observatory is a facility of the National Science Foundation operated under cooperative agreement by Associated Universities, Inc. The search for an optical counterpart was based on observations made with the NASA/ESA \emph{Hubble Space Telescope}, and obtained from the Hubble Legacy Archive, which is a collaboration between the Space Telescope Science Institute (STScI/NASA), the Space Telescope European Coordinating Facility (ST-ECF/ESA) and the Canadian Astronomy Data Centre (CADC/NRC/CSA). The scientific results reported in this article are based to a significant degree on data obtained from the Chandra Data Archive.\\

{\it Facilities:} \facility{VLA}, \facility{CXO}, \facility{HST}

\bibliography{gc.bib}

\end{document}